\documentclass{article}
\usepackage{spconf,amsmath,epsfig}
\usepackage{multirow}
\usepackage{hyperref}

\let\OLDthebibliography\thebibliography
\renewcommand\thebibliography[1]{
  \OLDthebibliography{#1}
  \setlength{\parskip}{0pt}
  \setlength{\itemsep}{0pt plus 0.3ex}
}

\begin{document}\sloppy

\def\x{{\mathbf x}}
\def\L{{\cal L}}

\title{Towards Proper Contrastive Self-supervised Learning Strategies For Music Audio Representation}
%



\name{
    Jeong Choi\textsuperscript{\rm 1}  \qquad
    Seongwon Jang\textsuperscript{\rm 2}\sthanks{Currently works at Tmap Mobility Co., Ltd.}
    \qquad
    Hyunsouk Cho\textsuperscript{\rm 3}\sthanks{Corresponding author.}
     \qquad
    Sehee Chung\textsuperscript{\rm 2}
}
\address{\textsuperscript{\rm 1}NAVER Corp.  \qquad  
         \textsuperscript{\rm 2}Knowledge AI Lab., NCSOFT  \qquad
         \textsuperscript{\rm 3}Ajou University 
         \\
         \textsuperscript{\rm 1}jeong.choi@navercorp.com;
         \textsuperscript{\rm 2}\{swjang90, seheechung\}@ncsoft.com;
         \textsuperscript{\rm 3}hyunsouk@ajou.ac.kr
         }


\maketitle

\begin{abstract}
The common research goal of self-supervised learning is to extract a general representation which an arbitrary downstream task would benefit from. In this work, we investigate music audio representation learned from different contrastive self-supervised learning schemes and empirically evaluate the embedded vectors on various music information retrieval (MIR) tasks where different levels of the music perception are concerned.
We analyze the results to discuss the proper direction of contrastive learning strategies for different MIR tasks.
We show that these representations convey a comprehensive information about the auditory characteristics of music in general, although each of the self-supervision strategies has its own effectiveness in certain aspect of information. 
\end{abstract}
\section{Introduction}\label{sec:introduction}
\label{sec:intro}
\subsection{Contrastive Self-supervised Learning}

Self-supervised learning has great potential in retrieving informative representation from a large amount of unlabeled data. Especially, deep learning architectures for this paradigm have been extensively studied recently in different research fields. 
Out of the major forms of self-supervision, one that we focus on is the contrastive learning approach that leverages a classification objective for differentiating positive and negative examples. Recent contrastive learning approaches have been especially successful on the general representational learning task in various domains with the emerging innovations on deep learning architectures.

The \textit{distributional similarity} between samples and the augmentation-invariant data characteristics are the two core elements of contrastive learning scheme in different domains that require careful designs of `\textit{pretext}' tasks \cite{misra2020self}. 
The \textit{distributional similarity} inherits the concept of distributed representations for words that have been hugely successful in the natural language processing field. It is achieved by predicting samples that are closely located or more probable to be within a same sequence under certain sequential context. This approach has been widely explored in the language domain \cite{mikolov2013distributed, peters2018deep, devlin2018bert} and adopted to the audio domain \cite{chung2016audio, Baevski2020wav2vec2A} and image domain \cite{chen2020generative}.
On the other hand, various effective data augmentation or transform (e.g. Fourier transform) techniques in different domains have been proposed for extraction of either augmentation-invariant information or the transformation objectives themselves. They have been actively studied for image \cite{noroozi2016unsupervised, zhang2016colorful, chen2020simple}, audio \cite{jansen2017unsupervised, pascual2019learning, ravanelli2020multitask}, or multimodal \cite{owens2018audio} self-supervision.


\vspace{-2mm}
\subsection{Self-supervised Learning of Audio Representation}

From the two types of self-supervision approaches, an approach leveraging \textit{distributional similarity}
was explored in \textit{CPC}, \textit{APC}, \textit{Audio ALBERT}, and \textit{Wav2vec 2.0} for audio representation. \cite{oord2018representation, chung2020apc, chi2021albert, Baevski2020wav2vec2A}. These models are trained to predict the future or masked segments from the input sequence. The learned embeddings are basically targeted to represent underlying coherent characteristic of a certain audio sequence that can discriminate itself from distractors \cite{oord2018representation}, which can be either other sequences or other parts from the same sequence by using a triplet or \textit{infoNCE} loss function. \textit{Vq-wav2vec} \cite{Baevski2020vq-wav2vec} and \textit{Wav2vec 2.0}  \cite{Baevski2020wav2vec2A} add quantization steps which force the model to represent an audio segment into a fixed number of discretized labels. This quantization steps along with the diversity loss function \cite{Baevski2020wav2vec2A} encourage the model to focus more on intra-sequence discrimination.
These models has been mainly studied towards speech related tasks. 
\textit{COLA} \cite{Saeed2021COLA} is an attempt to train a general purpose audio representation for various sound classification task. It does not take the sequential order into account, however, they chose input and target samples by randomly cropping from the same audio to extract coherent information in a single audio sequence.

On the other hand, the second group of audio self-supervision models are targeted to an objective of extracting augmentation-invariant features \cite{jansen2017unsupervised, pmlr-v130-al-tahan21a, Kharitonov2021}. By maximizing agreement between an audio segment and the augmented version of it, these models learn to keep the information that are not affected by the augmentation procedure. 
They are also trained with a triplet or \textit{infoNCE} loss function that leverage distractors to learn inter-sample discriminative features. 


The self-supervised embeddings are usually evaluated on a few different downstream tasks from the domain. 
In the speech domain, they are usually evaluated on two problems; the phoneme recognition and the speaker identification. These two tasks require exactly opposite notions of audio features. The phoneme recognition task would take advantage of speaker-invariant local characteristics of short audio segment, while the speaker identification task would require a phoneme-invariant global feature of a full speech sequence (e.g. timbre). \textit{Vq-wav2vec} and \textit{Wav2vec 2.0} have reached the state-of-the-art level of phoneme recognition scores, while \textit{CPC} and \textit{Audio ALBERT} show relatively poor performance. We argue that the performance gap is mainly caused from the quantization step, since the task requires intra-sequence discriminativeness. \footnote{\cite{Baevski2020wav2vec2A} conducted an ablation experiment related to this matter.} When it comes to the speaker identification task, however, \textit{CPC} and \textit{Audio ALBERT} also achieve the score almost as high as the supervised state-of-the-art. \cite{oord2018representation, chi2021albert, Baevski2020vq-wav2vec, Baevski2020wav2vec2A} 

Self-supervised audio representations are also evaluated on various sound classification tasks. Since these tasks are to classify the entire audio sequence into a single label, intra-sequence discrimination is not as important as in the phoneme recognition task. Most audio data augmentation based approaches are evaluated on these tasks. By carefully designing the augmentation procedures, these models achieved performances comparable to the supervised ones. \cite{jansen2017unsupervised, pmlr-v130-al-tahan21a, Kharitonov2021}


\vspace{-2mm}
\subsection{Music Audio Representation}

Representational learning aims to extract information that is useful for training wide range of classifiers or other predictors while being less specialized in a single supervised task \cite{oord2018representation,bengio2013representation}. In case of music audio, this leads to a question of what level of data characteristic each of different music-related classification tasks would demand. 

Defining a similarity metric between music audio data encompasses a wide range of perspectives from objective descriptors to human subjective perceptions. As a result, it is considered to be difficult to pinpoint the exact task-relevant information from music audio for an individual music information retrieval (MIR) task \cite{kim2020one}.
An early work \cite{slaney2008learning} had used mid-level information inferred from audio, user feedback data, and metadata to define multiple different Euclidean metric spaces of music similarity. Another work \cite{mcfee2012learning} had proposed the content-based music similarity metric by leveraging a sample of collaborative filter data along with the audio. There also was an in-depth investigation of the characteristics of multiple deep music representations learned from different supervised tasks along with the benefits from multi-task learning approaches \cite{kim2020one}. Other recent works \cite{learningtorank, lee2020disentangle} employed a triplet network approach on music audio using the similarity metric derived from semantic tag labels. 


A recent work \cite{spijkervet2021contrastive} leveraged the combination of various contrastive approaches to propose a self-supervised and data efficient learning method for the music auto-tagging task. Their proposed method, \textit{CLMR}, was evaluated on various auto-tagging datasets to show a comparable performance to supervised models. 

While the previous works on the music audio representation were targeted towards a single specific objective, our work focuses on assessing the potential of self-supervised music embeddings as a general representation.
Our attempts in this work do not aim at outperforming the state-of-the-art score for each MIR task. Instead, we set up experiments to compare the performance in various MIR tasks between different self-supervision strategies.
We investigate to what extent we can benefit from music audio representations learned from some of widely used contrastive learning schemes by analyzing the results on three different MIR tasks (instrument classification, genre classification, and music recommendation) which are considered to represent different aspects of music similarity.

Our experiments are set up using contrastive learning algorithms with variations in input / target instance sampling and model architectures, which are designed to capture different levels the music semantic - global or regional information. 
Our strategies are categorized in Table \ref{tab-strat}.

We then use the trained models as feature extractors and evaluate on different MIR tasks, where each task represents a certain abstraction level of information. We compare the self-supervised embeddings with MFCCs which has long been a solid baseline feature in audio classification tasks.


\begin{table}[h!]
\begin{center}
\begin{small}
\begin{tabular}{lccc}
& \textit{Local aspect} & $\longleftrightarrow$ & \textit{Global aspect} 
\\
\hline
\begin{tabular}{@{}l@{}}\textit{Input}  \\ \textit{audio} \end{tabular}
& 
Single segment 
&
-
&
\begin{tabular}{@{}c@{}}Sequence of\\ segments 
\end{tabular}
\\
\hline
\begin{tabular}{@{}l@{}}\textit{Target}  \\ \textit{audio} \end{tabular}
& 
\begin{tabular}{@{}c@{}}
Self-augmented
\\
segment
\end{tabular}
&
\begin{tabular}{@{}c@{}}
Adjacent
\\
segment
\\
(same track)
\end{tabular}
&
\begin{tabular}{@{}c@{}}
Random
\\
segment
\\
(same track)
\end{tabular}
\\

\hline
\begin{tabular}{@{}l@{}}\textit{Conv.} \\ \textit{output} \end{tabular}
& 
\begin{tabular}{@{}c@{}}Multi-level \\ outputs 
\end{tabular}
&
-
&
\begin{tabular}{@{}c@{}}Highest-level \\ output
\end{tabular}
\\[6pt]  

\end{tabular}
\caption{\label{tab-strat}Self-supervision strategies covered in this work}
\end{small}
\end{center}
\vskip -0.2in
\vspace{-6mm}
\end{table}

\vspace{-2mm}
\section{Methodologies}
\label{sec:model}

\vspace{-2mm}
\subsection{Audio Feature Encoder}
\vspace{-1mm}
For our multiple self-supervision settings, we use the same audio encoding architecture that takes a time-frequency domain representation (mel-spectrogram) as an input.

We build a standard audio 2D CNN architecture with 5 layers of 5$\times$5 convolutional kernels~\cite{choicrnn}. Batch normalization and 2D max-pooling are applied to the output of each convolutional layer. 
The encoder will output a single feature vector for every 3-second sample of mel-spectrogram input. 
The encoded vectors are then fed into the self-supervision phase where diverse architectures and objective functions are concerned. 
Details are provided in our code.\footnote{https://github.com/kunimi00/ContrastiveSSLMusicAudio}

\vspace{-2mm}
\subsection{Contrastive Learning Model Architectures}

\subsubsection{Siamese Network}
We first take a metric learning approach using a siamese network architecture. We experiment with two types of loss functions; an \textit{infoNCE} loss and a multi-level \textit{infoNCE} loss. 


The \textit{InfoNCE} loss \cite{oord2018representation} inherits the concept of noise contrastive estimation \cite{gutmann10nce} while computing the mutual information between the encoded representation vectors in order to capture the similarity in the high-level abstraction~\cite{hjelm2018learning}. 

Given an anchor audio $x^a$ with 1 positive segment $x^p$ and $N-1$ negative segments $\{x^n_1, . . . x^n_{N-1}\}$ along with an encoding function $f$, the \textit{InfoNCE} loss is computed as follows:
\begin{multline}
\mathcal{L}_\textit{InfoNCE} (y^a, y^p, \{y_1^n, ...y_{N-1}^n\})\\
= -\log \frac{\exp \left( y^{a^{T}} y^p\right)}
{\exp (y^{a^{T}} y^p) + \sum_{j=1}^{N-1} \exp (y^{a^{T}} y^{n_{j}})},
\end{multline}
where $y$ denotes the output from the embedding function $f$ given an input segment $x$ ($y=f(x)$). Although the \textit{InfoNCE} loss was originally proposed in \textit{CPC} architecture \cite{oord2018representation} which we will describe in the next section, we adopt it into the siamese network architecture for comparison. $N-1$ negative samples ($x^n$) are randomly sampled from the other tracks in the same minibatch for computational efficiency \cite{oord2018representation}.

Being inspired by previous works \cite{lee2017multi, jiang2020similarity}, we also formulate a multi-level convolutional output loss. We add a fully-connected layer on top of outputs from each of 5 convolutional layers of the audio encoder. \textit{InfoNCE} loss is then computed at each level separately, and later summed up for the final loss term. It can be defined as follows:
\begin{equation}
\mathcal{L}_{\textit{InfoNCE-Multi}} = \sum_{i}^M \mathcal{L}_\textit{InfoNCE}(\hat{y_i}^a, \hat{y_i}^p, \{\hat{y_i}_{{1}^n}, ... \hat{y_i}_{{N-1}^n}\}) \\
\end{equation}
where $\hat{y_i}$ denotes the output from $i^{\text{th}}$ convolutional layer of the encoder followed by an additional fully-connected layer, and $M$ denotes the number of layers in the CNN encoder. 

\vspace{-2mm}
\subsubsection{Contrastive Predictive Coding (CPC)}
The core idea of \textit{CPC} \cite{oord2018representation} is to learn high-level features that are coherent along the whole sequence. To do so, they defined a loss term using \textit{InfoNCE}, where the mutual information between the latent feature extracted from the present input sequence and the one from a future segment is maximized. 
It can also be interpreted as a metric learning architecture with additory use of sequential information of the input data over time. 
\textit{CPC} loss is formulated as follows:
\begin{equation}
\begin{split}
\mathcal{L}_\text{CPC} 
& = - \sum_{k} \log p(c_t \vert f(x_{t+k}), \{f(x_1^n), ...f(x_{N-1}^n)\}) \\
& = \sum_{k} \mathcal{L}_\textit{InfoNCE}(c_t, y_{t+k}, \{y_1^n, . . .y_{N-1}^n\}) \\
\end{split}
\end{equation}
where $c_t$ is the output from the last timestep $t$ of a sequential module (1-layered GRU), given a sequence of encoded vectors from the CNN module using $t$ consecutive audio segments. $x_{t+k}$ is a positive sample that is $k$ segments away from the $t$-th segment within the same track. Again, we sample $N$-1 negative samples ($x^n$) from other tracks in the same minibatch. We denote our \textit{CPC} model as MelCPC since we take the mel-spectrogram input unlike the original one.

\vspace{-2mm}
\subsection{Target Instance Strategies}

\subsubsection{Audio augmentation}
For audio augmentation, we use pitch shifting, time stretching, reverberation, noise addition, and polarity inversion\cite{Kharitonov2021}. We randomly choose from \{-2, -1, 1, 2\} semitones for pitch shifting, and stretch time with a speed factor randomly chosen from a range between 0.8 to 1.2. We apply reverberation and noise addition with a probability of 50\%, individually. 

\vspace{-2mm}
\subsubsection{Sampling strategies}
For the sampling strategy of the positive instances, we take two different options. One is taking an adjacent segment of the anchor segment and the other is taking a random segment from the same track. For the former, we left a 0.5 second gap between the neighboring segments to avoid `shortcuts' where the network simply learns to capture edge continuity~\cite{noroozi2016unsupervised}. For the latter one, we sample uniformly random segments from the rest of the track. As a music audio clip generally has very dynamic changes over sequence in auditory characteristics, an adjacent sample is more probable to be more similar to the input than a random sample from the entire sequence.
We take one positive sample and use $N-1$ other samples from the same training batch as negatives, as forementioned \cite{chen2020simple}.

\vspace{-2mm}
\section{Dataset}
\label{sec:data}
For the training of self-supervised models, we choose the subset of 0.2M tracks from MSD \cite{Bertin-Mahieux2011} which has been used as a benchmark split for training music auto-tagging models in previous works \cite{choicrnn, lee2017multi}. 
It is also the same subset used to train the supervised auto-tagging models that we compare with (Section \ref{sec:pretrained}). 

As for the audio input, we downsample each recording to 16 kHz and compute mel-spectrogram with 512-point Hanning window, 512-point FFT and 256-point hop. We standardize the input across all training data for each experiment. A segment of 188 frames (3s) is fed into the encoder for the siamese networks, and 752 frames (12s) for the MelCPC model.

\vspace{-3mm}
\section{Experiments}
\label{sec:exp}
\vspace{-2mm}
To first verify that our proposed model's performance is comparable to the state-of-the-art level, we evaluate it on the same benchmark test set of MSD where the previous state-of-the-art works on music auto-tagging have been evaluated on.

We then evaluate the self-supervised embeddings on three different downstream MIR tasks. 
We suppose that each of these tasks indicates measuring a different level of music audio similarity.
The genre classification task would require high-level understanding of comprehensive audio information, whereas the instrument classification task would benefit from low-level details that represent the timbral information. The music recommendation task deals with the most subjective and abstractive aspects of music audio among the three. 


Following a standard procedure for evaluating the representational power of a pre-trained embeddings \cite{chen2020simple}, all experiments are conducted in a transfer learning setting where model weights are fixed after being trained to function as a feature encoder. After encoding each segment of the input audio into an embedding, we summarize it into a concatenated vector of mean and standard deviation for each dimension. For the MelCPC model, we feed the entire sequence of segments to the trained model at once, and use the outputs from all timesteps of GRU module to obtain the summarized vector. 
We evaluated the summarized vectors using a support vector machine with a linear kernel and a linear logistic regression classifier \cite{zhang2017split}.
For downstream tasks, we also use MFCCs as input for the baseline experiment.

\vspace{-2mm}
\subsection{Comparison with State-of-the-art}
\label{sec:pretrained}
\vspace{-1mm}
Although our main objective is not about outperforming existing methodologies in a specific task, we still aim to verify that our models have comparable representational power to the state-of-the-arts in an arbitrary task. 
We compare one of our model, MelCPC, with two state-of-the-art fully-supervised models \cite{musicnn, won2021semi} and a recent self-supervised model \cite{spijkervet2021contrastive} on the music auto-tagging task. All models are trained and evaluated with the same benchmark split of MSD (201,680 training / 11,774 validation / 28,435 test samples) annotated with 50 tags. For self-supervised models, an additional linear logistic regression classifier is trained using the output from the pre-trained self-supervised model. 



\vspace{-2mm}
\subsection{Genre Classification}
\label{subsec:genre}
\vspace{-1mm}
We set up a genre classification experiment using FMA \textit{small} dataset \cite{fma_dataset} which contains 8,000 tracks annotated with 8 different genres. We use the provided official splits~\footnote{https://github.com/mdeff/fma}. 

\subsection{Instrument Classification}
 
For the instrument classification task, we use the training subset from IRMAS dataset following the setting from~\cite{kim2020one}. There are 6,705 multi-instrumental audio clips (3s long) annotated with a predominant instrument class. 
As our audio encoder will output a single feature vector for 3-second long inputs, 
we do not need a summarizing step. We train a support vector machine and a linear logistic regression classifier \cite{zhang2017split} using MFCC and the pre-trained embeddings for evaluation.

\subsection{Music Recommendation}
AOTM 2011 dataset~\cite{mcfee2012hypergraph} is used for the music recommendation task.
Tracks overlapping with the training set of our self-supervised models are excluded, along with playlists that contain too few or many tracks and ambiguous categories. 21,088 tracks in 7,245 playlists are remained from the original set.

We adopt the evaluation protocol used in~\cite{benzi2016song}. Given 3 query items (tracks) from each playlist, a recommender algorithm is asked to return the remaining ground-truth items. We run ItemKNN-CBF~\cite{lops2011content, dacrema2019we} using cosine distance function to predict the rankings of the ground-truth items.
To avoid heavy computation on all item-item pairs, we follow the evaluation scheme from ~\cite{he2017neural} by pairing each ground-truth item with 100 random negative items. Hit ratio@10 (\textit{HR@10}) and mean percentile rank (\textit{MPR}) are measured from the ranked lists. \textit{HR@10} computes the fraction of times that the ground-truth items are among the top 10 returned items, and \textit{MPR} is a position-aware metric that assigns larger weights to higher positions (i.e., $1/i$ for the $i^{\text{th}}$ position in the ranked list).

\begin{table}[t!]
\begin{center}
\begin{tabular}{ccc}
\textit{Supervision} & \textit{Model} & \textit{AUC-ROC}  \\ 
\hline
\hline
supervised & Transformer \cite{won2021semi} & 0.897 \\
supervised & musicnn \cite{musicnn} & 0.880 \\
\hline
self-supervised & CLMR \cite{spijkervet2021contrastive} & 0.857 \\ 
self-supervised & MelCPC (ours) & \textbf{0.856} \\ 
\end{tabular}
\caption{\label{exp-msdtest} Auto-tagging results on \textit{MSD} benchmark subset.}
\end{center}
\vskip -0.2in
\vspace{-2mm}
\end{table}


\begin{table*}[]
\centering
\begin{tabular}{llccccccccc}
\multicolumn{1}{c}{\multirow{2}{*}{Feature}} & \multicolumn{1}{c}{\multirow{2}{*}{Loss}} & \multirow{2}{*}{Dim.} & \multicolumn{2}{c}{Genre classification} & \multicolumn{1}{l}{} & \multicolumn{2}{c}{Inst. classification} & \multicolumn{1}{l}{} & \multicolumn{2}{c}{Music rec.} \\ \cline{4-5} \cline{7-8} \cline{10-11} 
\multicolumn{1}{c}{}                         & \multicolumn{1}{c}{}                      &                       & \textit{SVM}                 & \textit{LR}                 &                      & \textit{SVM}                   & \textit{LR}                    &                      & \textit{HR@10}               & \textit{MPR}                \\ \hline
MFCC                                         & \multicolumn{1}{c}{-}                     & 40                    & 0.397               & 0.401              &                      & 0.473                 & 0.481                 &                      & 0.196               & 0.381              \\ \hline
Siamese-\textit{Self-aug}                             & \textit{InfoNCE}                                   & 128                   & 0.409               & 0.418              &                      & 0.431                 & 0.438                 &                      & 0.250               & 0.339              \\
Siamese-\textit{Self-aug}                             & \textit{InfoNCE-Multi}                             & 640                   & 0.373               & 0.413              &                      & 0.520                 & 0.532                 &                      & 0.242               & 0.350              \\ \hline
Siamese-\textit{Adjacent}                             & \textit{InfoNCE}                                   & 128                   & 0.466               & 0.458              &                      & 0.388                 & 0.379                 &                      & 0.279               & 0.318              \\
Siamese-\textit{Adjacent}                             & \textit{InfoNCE-Multi}                             & 640                   & 0.450               & 0.446              &                      & \textbf{0.592}        & \textbf{0.575}        &                      & 0.262               & 0.332              \\ \hline
Siamese-\textit{Random}                               & \textit{InfoNCE}                                   & 128                   & 0.418               & 0.431              &                      & 0.374                 & 0.361                 &                      & 0.239               & 0.377              \\
Siamese-\textit{Random}                               & \textit{InfoNCE-Multi}                             & 640                   & 0.436               & 0.461              &                      & \textbf{0.581}        & \textbf{0.560}        &                      & 0.260               & 0.330              \\ \hline
MelCPC                                          & \textit{InfoNCE}                                   & 256                   & \textbf{0.480}      & \textbf{0.491}     &                      & 0.490                 & 0.486                 &                      & \textbf{0.280}      & \textbf{0.316}    
\end{tabular}
\caption{\label{exp-all} Results on three downstream tasks.
(\textit{Self-aug} : self-augmentation approach /  \textit{Adjacent} : positive sampling of an adjacent segment / \textit{Random} : positive sampling from the entire track /  \textit{InfoNCE-Multi} : \textit{InfoNCE} loss aggregated from multi-level convolutional layer outputs / 
\textit{SVM} : accuracy using a support vector machine classifier with a linear kernel / \textit{LR} : a linear logistic regression classifier \cite{zhang2017split} /  \textit{HR@10} : hit ratio at 10 (higher the better) / \textit{MPR} : mean percentile rank (lower the better).}

\vskip -0.1in
\vspace{-1mm}
\end{table*}

\section{Results and Discussion}
\label{sec:result}

\subsection{Comparison with State-of-the-art}
From Table \ref{exp-msdtest}, we can verify that our model has comparable representational power even with the state-of-the-art level models \cite{musicnn, won2021semi} that are trained with the same audio set in a fully-supervised manner. It also shows a similar performance with a recently proposed self-supervised music embedding model that employs the combination of various contrastive learning techniques on audio \cite{spijkervet2021contrastive}. 

\subsection{Genre Classification Results}

Table \ref{exp-all} shows experimental results for the genre classification problem. As the genre classification task relies on high level understanding of music audio, MelCPC employing sequential information to better summarize higher level data abstraction over time performs better than siamese networks that only take singular segments to be compared. 
It is also interesting to see that the multi-layer output model performs worse than a single-layer output model for adjacent sampling-based and augmentation-based models, indicating that the lower level features are not helping when it comes to a problem of high level music understanding in those cases.

However, when sampling targets randomly from the entire track, multi-layer output model outperforms. We suspect that this is because, in case of leveraging more similar audio segments as input and target (adjacent sampling and augmentation), features extracted from low level convolutional layers can be redundant, while more informative features can be extracted when less similar input and target segments are sampled from the entire track. We find similar trend in music recommendation task also. 
\vspace{-2mm}

\subsection{Instrument Classification Results}

The instrument classification results are also shown in Table \ref{exp-all}.
In this task, multi-layer output models outperform single output ones in all cases. When employing single-layer output, data-augmentation approach shows the highest score among all siamese networks. However, using a multi-layer output model for sampling approaches increased the performance in a greater deal, resulting the best performing model to be the one with adjacent target sampling approach. This indicates that, for the data augmentation approach, adding multi-layer outputs is not as helpful as in in-track sampling approaches because audio augmentation already concerns low-level information to some degree. MelCPC performed better than all single output siamese networks, but still poorer than multi-level output ones. 

\vspace{-3mm}
\subsection{Music Recommendation Results}

As shown in Table \ref{exp-all}, recommendation task results show very similar trend to the genre classification results. 
The MelCPC model shows the best performance, and we suspect that this is because the recommendation task deals with rather implicative and subjective level of information. It is not easy to define what level of auditory perception is concerned with regard to music recommendation, but we can induce from the results that higher level information is more related.

\vspace{-2mm}
\subsection{Target Instance Strategies}

Regarding the target instance strategies overall, sampling an positive segment from the same audio track gives a better cue for the genre classification and the recommendation task than for the instrument classification task. When sampling, choosing an adjacent segment was more effective than randomly picking from the entire sequence for all cases.

\vspace{-2mm}
\section{Conclusion and Future Work}
\label{sec:future}
\vspace{-1mm}

In this work, we explore diverse directions of self-supervision strategies for different MIR tasks. 
We verify that, since MIR tasks cover wide range of auditory characteristics and are generally a more subjective matter compared to other audio domains, it is important to carefully choose right strategies of self-supervision for a certain desired task. 


For future works, MIR tasks that require intra-sequence discriminative representation, such as music transcription or chord recognition, can further be considered.
As complex as MIR tasks are compared to speech or general sound classification problems, novel self-supervision architectures or
\textit{pretext} tasks \cite{misra2020self} specially targeted to music audio analysis might contribute to some breakthroughs.

\vspace{-2mm}
\section{Acknowledgement}
\label{sec:acknowledgement}
\vspace{-1mm}
We thank Keunwoo Choi for helpful review and discussion.



\vspace{-2mm}
\bibliographystyle{IEEEbib}
\bibliography{icme2022SSLmusic}

\begin{thebibliography}{10}

\bibitem{misra2020self}
I.~Misra et~al.,
\newblock ``Self-supervised learning of pretext-invariant representations,''
\newblock in {\em CVPR Workshops}, 2020.

\bibitem{mikolov2013distributed}
T.~Mikolov et~al.,
\newblock ``Distributed representations of words and phrases and their
  compositionality,''
\newblock in {\em NeurIPS}, 2013.

\bibitem{peters2018deep}
M.~Peters et~al.,
\newblock ``Deep contextualized word representations,''
\newblock in {\em NAACL}, 2018.

\bibitem{devlin2018bert}
J.~{Devlin} et~al.,
\newblock ``Bert: Pre-training of deep bidirectional transformers for language
  understanding,''
\newblock in {\em NAACL}, 2019.

\bibitem{chung2016audio}
Y.~{Chung} et~al.,
\newblock ``Audio word2vec: Unsupervised learning of audio segment
  representations using sequence-to-sequence autoencoder,''
\newblock {\em CoRR}, 2016.

\bibitem{Baevski2020wav2vec2A}
A.~{Baevski} et~al.,
\newblock ``wav2vec 2.0: A framework for self-supervised learning of speech
  representations,''
\newblock {\em ArXiv}, 2020.

\bibitem{chen2020generative}
Mark Chen et~al.,
\newblock ``Generative pretraining from pixels,''
\newblock in {\em ICML}, 2020.

\bibitem{noroozi2016unsupervised}
M.~Noroozi et~al.,
\newblock ``Unsupervised learning of visual representations by solving jigsaw
  puzzles,''
\newblock in {\em ECCV}, 2016.

\bibitem{zhang2016colorful}
R.~Zhang et~al.,
\newblock ``Colorful image colorization,''
\newblock in {\em ECCV}, 2016.

\bibitem{chen2020simple}
T.~Chen et~al.,
\newblock ``A simple framework for contrastive learning of visual
  representations,''
\newblock {\em ICML}, 2020.

\bibitem{jansen2017unsupervised}
A.~{Jansen} et~al.,
\newblock ``Unsupervised learning of semantic audio representations,''
\newblock in {\em ICASSP}, 2018.

\bibitem{pascual2019learning}
Santiago P. et~al.,
\newblock ``{Learning Problem-Agnostic Speech Representations from Multiple
  Self-Supervised Tasks},''
\newblock in {\em Interspeech}, 2019.

\bibitem{ravanelli2020multitask}
M.~{Ravanelli} et~al.,
\newblock ``Multi-task self-supervised learning for robust speech
  recognition,''
\newblock in {\em ICASSP}, 2020.

\bibitem{owens2018audio}
Andrew Owens and Alexei~A Efros,
\newblock ``Audio-visual scene analysis with self-supervised multisensory
  features,''
\newblock in {\em ECCV}, 2018.

\bibitem{oord2018representation}
A.~Oord et~al.,
\newblock ``Representation learning with contrastive predictive coding,''
\newblock {\em ArXiv}, 2018.

\bibitem{chung2020apc}
Yu-An Chung et~al.,
\newblock ``Generative pre-training for speech with autoregressive predictive
  coding,''
\newblock in {\em ICASSP}, 2020.

\bibitem{chi2021albert}
Po-Han Chi et~al.,
\newblock ``Audio albert: A lite bert for self-supervised learning of audio
  representation,''
\newblock in {\em IEEE Spoken Language Technology Workshop}, 2021.

\bibitem{Baevski2020vq-wav2vec}
Alexei Baevski et~al.,
\newblock ``vq-wav2vec: Self-supervised learning of discrete speech
  representations,''
\newblock in {\em ICLR}, 2020.

\bibitem{Saeed2021COLA}
Aaqib Saeed et~al.,
\newblock ``Contrastive learning of general-purpose audio representations,''
\newblock in {\em ICASSP}, 2021.

\bibitem{pmlr-v130-al-tahan21a}
Haider Al-Tahan et~al.,
\newblock ``Clar: Contrastive learning of auditory representations,''
\newblock in {\em AISTATS}, 2021.

\bibitem{Kharitonov2021}
Eugene Kharitonov et~al.,
\newblock ``Data augmenting contrastive learning of speech representations in
  the time domain,''
\newblock in {\em IEEE Spoken Language Technology Workshop}, 2021.

\bibitem{bengio2013representation}
Y.~Bengio et~al.,
\newblock ``Representation learning: A review and new perspectives,''
\newblock {\em IEEE Transactions on Pattern Analysis and Machine Intelligence},
  2013.

\bibitem{kim2020one}
J.~Kim et~al.,
\newblock ``One deep music representation to rule them all? a comparative
  analysis of different representation learning strategies,''
\newblock {\em Neural. Comput. Appl.}, 2020.

\bibitem{slaney2008learning}
M.~{Slaney} et~al.,
\newblock ``Learning a metric for music similarity,''
\newblock in {\em ISMIR}, 2008.

\bibitem{mcfee2012learning}
B.~{McFee} et~al.,
\newblock ``Learning content similarity for music recommendation,''
\newblock {\em IEEE Trans Audio Speech Lang Process}, 2012.

\bibitem{learningtorank}
L.~{Prétet} et~al.,
\newblock ``Learning to rank music tracks using triplet loss,''
\newblock in {\em ICASSP}, 2020.

\bibitem{lee2020disentangle}
J.~{Lee} et~al.,
\newblock ``Disentangled multidimensional metric learning for music
  similarity,''
\newblock in {\em ICASSP}, 2020.

\bibitem{spijkervet2021contrastive}
Janne Spijkervet et~al.,
\newblock ``Contrastive learning of musical representations,''
\newblock in {\em ISMIR}, 2021.

\bibitem{choicrnn}
K.~{Choi} et~al.,
\newblock ``Convolutional recurrent neural networks for music classification,''
\newblock in {\em ICASSP}, 2017.

\bibitem{gutmann10nce}
M.~{Gutmann} et~al.,
\newblock ``Noise-contrastive estimation: A new estimation principle for
  unnormalized statistical models,''
\newblock in {\em JMLR}, 2010.

\bibitem{hjelm2018learning}
R~D. {Hjelm} et~al.,
\newblock ``Learning deep representations by mutual information estimation and
  maximization,''
\newblock in {\em ICLR}, 2019.

\bibitem{lee2017multi}
J.~Lee et~al.,
\newblock ``Multi-level and multi-scale feature aggregation using pretrained
  convolutional neural networks for music auto-tagging,''
\newblock {\em IEEE Signal Process Letters}, 2017.

\bibitem{jiang2020similarity}
C.~Jiang et~al.,
\newblock ``Similarity learning for cover song identification using
  cross-similarity matrices of multi-level deep sequences,''
\newblock in {\em ICASSP}, 2020.

\bibitem{Bertin-Mahieux2011}
T.~{Bertin-Mahieux} et~al.,
\newblock ``The million song dataset,''
\newblock in {\em ISMIR}, 2011.

\bibitem{zhang2017split}
R.~Zhang et~al.,
\newblock ``Split-brain autoencoders: Unsupervised learning by cross-channel
  prediction,''
\newblock in {\em CVPR}, 2017.

\bibitem{musicnn}
J.~{Pons} and X.~{Serra},
\newblock ``musicnn: Pre-trained convolutional neural networks for music audio
  tagging,''
\newblock {\em ISMIR Late Breaking Demo}, 2019.

\bibitem{won2021semi}
Minz Won, , et~al.,
\newblock ``Semi-supervised music tagging transformer,''
\newblock in {\em ISMIR}, 2021.

\bibitem{fma_dataset}
Micha\"el Defferrard et~al.,
\newblock ``{FMA}: A dataset for music analysis,''
\newblock in {\em ISMIR}, 2017.

\bibitem{mcfee2012hypergraph}
B.~McFee et~al.,
\newblock ``Hypergraph models of playlist dialects.,''
\newblock in {\em ISMIR}, 2012.

\bibitem{benzi2016song}
K.~Benzi et~al.,
\newblock ``Song recommendation with non-negative matrix factorization and
  graph total variation,''
\newblock in {\em ICASSP}, 2016.

\bibitem{lops2011content}
P.~Lops et~al.,
\newblock ``Content-based recommender systems: State of the art and trends,''
\newblock in {\em Recommender systems handbook}. Springer, 2011.

\bibitem{dacrema2019we}
M.~Dacrema et~al.,
\newblock ``Are we really making much progress? a worrying analysis of recent
  neural recommendation approaches,''
\newblock in {\em RecSys}, 2019.

\bibitem{he2017neural}
X.~He et~al.,
\newblock ``Neural collaborative filtering,''
\newblock in {\em WWW}, 2017.

\end{thebibliography}

\end{document}